%% file: quantified_permissions.tex
\documentclass[orivec]{llncs}
\pdfoutput=1 %% Required by arXiv to ensure processing with pdflatex

\usepackage[fleqn]{amsmath}
\usepackage{amssymb}
\usepackage{xfrac}
\usepackage{microtype}
\usepackage{xspace}
\usepackage{mathabx}
\usepackage{algorithm2e}
\usepackage{silver}
\usepackage{suffix}
\usepackage[T1]{fontenc}
\usepackage{lmodern}
\usepackage{textcomp}
\IfFileExists{luximono.sty}{\usepackage[scaled=0.9]{luximono}}{\usepackage[scaled=0.81]{beramono}}
\usepackage{afterpage}
\usepackage{dcolumn}
\usepackage{makecell}
\input{macros.tex}
\begin{document}

\title{Automatic Verification of Iterated Separating Conjunctions using
Symbolic Execution}

\author{Peter M\"uller \and Malte Schwerhoff \and Alexander J. Summers
}
\institute{
Department of Computer Science, ETH Zurich, Switzerland \\
\email{\{peter.mueller, malte.schwerhoff, alexander.summers\}@inf.ethz.ch}
%\email{peter.mueller@inf.ethz.ch} \email{malte.schwerhoff@inf.ethz.ch} \email{alexander.summers@inf.ethz.ch}
}

\maketitle
\pagestyle{plain}

\begin{abstract}
In permission logics such as separation logic, the iterated separating
conjunction is a quantifier denoting access permission to an unbounded
set of heap locations. In contrast to recursive
predicates, iterated separating conjunctions do not prescribe a
structure on the locations they range over, and so do not
restrict how to traverse and modify these locations. This flexibility is
important for the verification of random-access data structures such
as arrays and data structures that can be traversed in multiple ways
such as graphs. Despite its usefulness, no automatic program verifier
natively supports iterated separating conjunctions; they are especially difficult to incorporate into symbolic execution engines, the prevalent technique for
building verifiers for these logics.

In this paper, we present the first symbolic execution technique to
support general iterated separating conjunctions.  We propose a novel representation of symbolic heaps and flexible support
for logical specifications that quantify over heap locations.  Our
technique exhibits predictable and fast performance despite
employing quantifiers at the SMT level, by carefully controlling
quantifier instantiations. It is compatible with other
features of permission logics such as fractional permissions,
recursive predicates, and abstraction functions. Our technique is implemented as an extension of the Viper verification infrastructure.

\end{abstract}

%%%%%%%%%%%%%%%%%%%%%%%%%%%%%%%%%%%%%%%%%%%%%%%%%%%%%%%%%%%%%%%%%%%%%
\section{Introduction}
\label{sec:introduction}
%%%%%%%%%%%%%%%%%%%%%%%%%%%%%%%%%%%%%%%%%%%%%%%%%%%%%%%%%%%%%%%%%%%%%

Permission logics such as separation logic \cite{Reynolds02a} and
implicit dynamic frames \cite{SmansEA09} associate an access
permission with each memory location in order to reason about shared
mutable state.  Dynamic heap data structures require specifications to
denote access permissions to a statically-unknown set of locations.
Such specifications are typically expressed in existing tools using recursive
predicates \cite{ParkinsonBierman05}, which work well
so long as the traversal of the data structure matches the definition of the
predicate. However, access patterns that do not follow the predicate
structure (\eg{}, traversing a doubly-linked list from the end)
or that follow no specific order (\eg{}, random access into an array)
are difficult to handle in existing program verifiers, requiring programmers to provide substantial manual proof steps
 (for instance, as ghost code) to bridge the mismatch between
the program's access pattern and the imposed predicate structure.

Iterated separating conjunction \cite{Reynolds02a} (hereafter, ISC) is an alternative
way to denote properties of a set of heap locations, which has
for instance been used in by-hand proofs to denote locations of arrays
\cite{Reynolds02a}, cyclic data structures
\cite{BirkedalTR04,Yang01anexample}, the objects stored in
linked lists \cite{Distefano:2008}, and graph algorithms \cite{Yang01anexample}.  Unlike recursive predicates, an
\isc\ does not prescribe any particular traversal order.

% Bart \cite{jacobs2015} uses ISC to denote sets of call permissions.

Despite its usefulness and inclusion in early presentations of separation logic, no existing program verifier supports general
\iscs\ directly.  Among the tools based on symbolic execution,
Smallfoot~\cite{becaoh2006} does not support \isc; VeriFast~\cite{SmansJP13} and jStar \cite{Distefano:2008} allow programmers to encode
some forms of \isc\ via abstract predicates that can be manipulated by
auxiliary operations and lemmas (in VeriFast) or tailored rewrite
rules (in jStar). For arrays,
this encoding is partially supported by libraries.
However, in the general case,
programmers need to provide the extra machinery, which significantly increases the
necessary manual effort.

Among the verifiers based on verification condition generation,
Chalice \cite{LeinoMueller09} supports only a restricted form of \isc{}
(ranging over all objects stored in a sequence), and
VeriCool uses an encoding
that leads to unreliable behaviour of the SMT solver~\cite[p.~46]{Smans:2012}. The GRASShopper tool \cite{PiskacWZ14} does not provide built-in or general support for \isc{}, but some ingredients of the technique we present (particularly, the technical usage of inverse functions) have been employed there to specify particular random access to data structures (\eg, arrays). The Dafny verifier \cite{Leino:2010:DAP:1939141.1939161} can be used to write similar set and quantifier-based specifications, but does not support permission-based reasoning or concurrency.

In this paper, we present the first symbolic execution technique that
directly supports general forms of \isc. Our technique is compatible
with other features of permission logics: it supports
fractional permissions \cite{Boyland}, such that a heap
location may be ranged over by several \iscs, and allows \isc\ to
occur in predicate bodies and in preconditions of abstraction
functions \cite{HeuleKassiosMuellerSummers13}.

This combination of features allows one to specify and verify
%advances the state-of-the-art of automated
%verification of allows the specification and verification of
%complex functional properties of
challenging examples such as
graph-marking algorithms (see \appref{appendix_graph}) that so far were beyond the scope of automated
verifiers based on permission logics.

Our main technical contributions are:
(1)~a novel representation of the partial heaps that are denoted by an
\isc, along with algorithms to manipulate this representation;  (2)~a
technique to preserve across heap changes (to frame) the values of expressions
that depend on the  unbounded set of heap locations
denoted by \iscs; (3)~an SMT encoding that carefully
controls quantifier instantiations; (4)~an implementation of our approach in the Viper verification
infrastructure~\cite{MuellerSchwerhoffSummers16}.
Our implementation and several interesting examples are available 
online \cite{ViperOnline}.

\subsubsection{Outline.} In the next section, we explain the
main technical challenges our work addresses, and illustrate them with
a simple motivating example. Our design for a symbolic heap that can represent permissions described by \iscs{} is presented
in \secref{representation}. We explain the symbolic evaluation of
expressions and framing with respect to this heap representation in \secref{symbolic_values}.
In \secref{quantifiers}, we discuss how we control quantifier instantiations.
\secref{implementation} presents an evaluation of our implementation. We conclude in \secref{conclusion}.

%%%%%%%%%%%%%%%%%%%%%%%%%%%%%%%%%%%%%%%%%%%%%%%%%%%%%%%%%%%%%%%%%%%%%
\section{Technical Challenges}
\label{sec:background}
%%%%%%%%%%%%%%%%%%%%%%%%%%%%%%%%%%%%%%%%%%%%%%%%%%%%%%%%%%%%%%%%%%%%%

Permission logics ensure that a heap location is accessed only
  when the corresponding permission is held. Dedicated assertions
  denote the permission to a heap location $e.f$, written as
  $\pointsto{e.f}{\_}$ in separation logic and as the \emph{accessibility
    predicate} \inlinesilver+acc($e.f$)+ in implicit dynamic frames; we use
  the latter in this paper. These logics include a \emph{separating
  conjunction} \inlinesilver+$\ast$+, expressing that the permissions
denoted by the two conjuncts must be disjoint.
For instance, an assertion
  \inlinesilver+acc($x$.$f$) $\ast$ acc($y$.$f$)+ implies the disequality $x\neq y$.
Many permission logics allow permissions to be split into fractions, and
to re-assemble fractions into a full permission. In these logics,
any non-zero permission allows read access to a location, whereas
write access requires the full permission.
 When appropriate permissions are held, assertions may also
  constrain the value of a heap location (for instance, $x.f > 3$);
  assertions that do not contain accessibility predicates are called
  \emph{pure}. We use the terms \emph{pure assertion} and
\emph{expression} synonymously.

 Verification of many program constructs can be modelled by two
  basic operations.  \term{Inhaling} an assertion $A$ adds the
  permissions denoted by $A$ to the current state and assumes the pure
  assertions in $A$.
  \term{Exhaling} an assertion $A$ checks that
  the current state satisfies the pure assertions in $A$; it also
  checks that the state contains the permissions denoted by $A$ and
  removes them. As soon as permission to a heap location is no longer
  held, information about its value cannot %can no longer
  be retained.  Inhale
  and exhale can be seen as the permission-aware analogues of assume
  and assert statements \cite{LeinoMueller09}; they are sometimes
  called produce and consume \cite{SmansJP10}. Using these operations,
  a method call (for example) can be encoded by exhaling the method precondition and
  then inhaling its postcondition.

Building a verification tool for a permission logic
requires effective solutions to the following \emph{technical challenges}:
\begin{enumerate}
\item How to model the program state, including permissions and values?

\item How to check for a permission in a state?

\item How to add and remove permissions to and from a state?

\item How to evaluate (heap-dependent) expressions in a state?

\item When to preserve (frame) an expression's value
across heap changes?
\end{enumerate}
In the remainder of this section, we summarize how existing
verifiers solve these challenges for logics without \isc and then
explain how providing support for \isc complicates these challenges.

%%--------------------------------------------------------------------
\subsection{Smallfoot-style Symbolic Execution}
\label{sec:smallfoot}
%%--------------------------------------------------------------------

Smallfoot \cite{becaoh2006} introduced a symbolic execution
  technique that has become the state-of-the-art way of building verifiers for permission logics. It provides
  simple and efficient solutions to the technical challenges above:
(1)~A symbolic state consists of a set of heap chunks, and a
set of path conditions. A \emph{heap chunk} has the form
$\pointsto{o.f}[v,p]$, mimicking separation logic's points-to predicates.
It records a receiver value $o$, a field name $f$,
a \emph{location value} $v$ representing the value stored in location
$o.f$, and a permission amount $p$. A permission amount is a
  value between 0 and 1 (inclusive); intermediate values can be
    used to support fractional permissions.
Here, $o$, $v$, and $p$ are (immutable) symbolic values.
\emph{Path conditions} are boolean constraints on the symbolic values collected
while verifying a program path such as the 
branch conditions on that path.
%They include user specifications and branch conditions. 
Path conditions may constrain heap values and may be quantified.
An SMT solver is used to
answer queries about the path conditions, for instance, equality of symbolic
values.
(2)~Checking for permission to a heap location entails iterating through
the heap chunks and finding those with matching receiver-field pairs.
(3)~Removing permissions is modelled by subtracting permissions from the
corresponding chunk(s), and adding a permission is modelled by adding a
heap chunk (with a fresh symbolic location value) that provides the added
permission amount.
(4)~Evaluating a heap lookup \inlinesilver+$e$.$f$+ yields the location value
of the chunk for $e.f$ (and is not permitted if no such chunk exists).
(5)~Framing the value of such expressions happens implicitly so long as the
same heap chunk provides non-zero permission to the location.
When a chunk no longer provides any permission, 
it gets removed and its location value becomes inaccessible.

In order to specify unbounded heap structures, the Smallfoot approach
has been extended to handle user-defined recursive predicates.  In
successor tools such as VeriFast \cite{SmansJP13}, jStar
\cite{Distefano:2008}, and
  Viper~\cite{MuellerSchwerhoffSummers16}, heap chunks may also
  represent predicate instances. Smallfoot-style symbolic execution
has also been extended to support heap-dependent pure
functions in the assertion language \cite{SmansJP10}. For example, the operations of a list class may be specified in terms
  of an \inlinesilver{itemAt} function. Such functions include a precondition
that requires permission to all locations read by the function body;
this information is used to frame function applications.

These extensions increase the expressiveness
of permission logics significantly, but are not sufficient
to simply specify and automatically reason about
important data structures such as arrays and graphs:
this requires support for \iscs{}.

%\as{\emph{Snapshots}} of the part of the
%  state on which a function's precondition depends \as{(represented as a tuple of one symbolic value per required access permission)} are used to frame
%  function values: in particular, when the verifier cannot evaluate
%  the function body due to information hiding or recursion.

%%--------------------------------------------------------------------
 \subsection{Iterated Separating Conjunction}
%%--------------------------------------------------------------------

\begin{figure}[t]
\begin{silver}[mathescape=true]
method Replace(a: Int[], left: Int, right: Int, from: Int, to: Int)
  requires 0 <= left < right <= a.length
  requires forall i: Int :: left <= i < right ==> acc(a[i])
  ensures  forall i: Int :: left <= i < right ==> acc(a[i])
  ensures  forall i: Int :: left <= i < right ==>
               (old(a[i]) == from ? a[i] == to : a[i] == old(a[i]))
{
  if (right - left <= 1) {
    if(a[left] == from) { a[left] := to }
  } else {
    var mid := left + (right - left) / 2
    Replace(a, left, mid, from, to)@{\hspace{3pt}\raisebox{-2pt}{\Large $\parallel$}\hspace{3pt}}@Replace(a, mid, right, from, to)
  }
}
\end{silver}
\vspace{-0.75em}
\caption{A parallel replace operation on array segments. The second
precondition and the first postcondition denote access permissions
to the elements of the array. The \inlinesilver{forall} quantifier in
these conditions denotes an \isc{}: the body of the quantifier
includes accessibility predicates (of the form \inlinesilver{acc(a[i])}).
The second postcondition uses a regular (pure) quantifier to specify the
functional behaviour of the method. Here,
\inlinesilver{old} expressions let the postcondition refer to values
in the prestate; the access permissions for these expressions come from
the second precondition.}
\label{fig:motivation}
%\vspace{-1.0em}
\end{figure}

\figref{motivation} illustrates the usage of \iscs{}: method
\inlinesilver{Replace} replaces all occurrences of integer
\inlinesilver{from} by integer \inlinesilver{to} in the segment of
array \inlinesilver{a} between \inlinesilver{left} and
\inlinesilver{right}. The recursive calls to smaller array segments
are performed concurrently using
parallel composition $\parallel$. The
second precondition requires access permissions for all elements in
the array segment, and the first postcondition returns these
permissions to the caller; both are expressed using \isc. The second postcondition specifies the
functional behaviour of the method using an
\inlinesilver{old}-expression to refer to the prestate of a method;
this pure assertion needs heap-dependent expressions under a
quantifier.

Verifying the example entails splitting the
symbolic state described by the \isc\ in the precondition in
  order to exhale the preconditions of the recursive calls, and to
re-combine the states resulting from inhaling the
postconditions of these calls after the parallel
composition, in order to prove the
callee's postcondition.

Providing support for \iscs complicates each of
  the five technical challenges discussed above:

\begin{enumerate}
\item Heap chunks must be generalised to denote permission to an
  unbounded number of locations simultaneously, and encode a symbolic
  value per location (for instance, to represent the values of each
  array location in \figref{motivation}).

\item Exhaling an \isc requires checking permission for an
  unbounded number of heap locations; these could be spread across
  multiple heap chunks, as in the case of exhaling the postcondition of \inlinesilver{Replace}.

\item Removing permissions from a generalised chunk may affect
  only some of the locations to which it provides permission. For example, when exhaling the
  precondition of the first recursive call to \inlinesilver{Replace},
  the permissions required for the second call must be retained in the symbolic
  state.

\item Evaluating heap-dependent expressions under quantifiers may rely
  on symbolic values from multiple heap chunks. For
  example, proving the second postcondition of \inlinesilver{Replace}
  requires information from both recursive
  calls.

\item
Framing in existing Smallfoot-style verifiers requires
that heap-dependent expressions depend only on a bounded number of
symbolic values (which can include representations of predicate instances~\cite{SmansJP10}). However, this requirement is too strong for pure quantifiers
over heap locations and for
functions whose preconditions use \iscs to require access to an unbounded
set of locations (see for instance the client of our running example,
either online \cite{ViperOnline} or in \appref{appendix_replace}).
\end{enumerate}

\noindent
Our technique is the first to provide automatic solutions to these challenging
problems. \secref{representation} tackles the first 3 problems; \secref{symbolic_values}
 tackles the remaining 2.

%%%%%%%%%%%%%%%%%%%%%%%%%%%%%%%%%%%%%%%%%%%%%%%%%%%%%%%%%%%%%%%%%%%%%
\section{Treatment of Permissions}
\label{sec:representation}
%%%%%%%%%%%%%%%%%%%%%%%%%%%%%%%%%%%%%%%%%%%%%%%%%%%%%%%%%%%%%%%%%%%%%
We consider the following canonical form of source-level
  assertion for denoting an \isc:
\qpforall{$x\colon T$}{$c(x)$}{acc($e(x).f$,$p(x)$)}, in which
$c(x)$ is a boolean
  expression, $e(x)$ a reference-typed expression, and $p(x)$ an
  expression denoting a permission amount.
More complex
    assertions can be desugared into this canonical form, for instance, iterating over the conjunction of two accessibility predicates can be encoded
    by repeating the quantification over each conjunct. For simplicity, we do not consider nested \iscs, but an extension
is possible.
Our canonical form is sufficient
  to directly model quantifying over receivers in a set (useful
  for graph examples, see \appref{appendix_graph}) or over integer indices into an array,
as shown in \figref{motivation}.

The permission expression
    $p(x)$ may be a complex expression including conditionals, and
    need not evaluate to the same value for each instantiation of
    $x$. This enables us to model complex access patterns such as
    requiring non-zero permission to every $n$th slot of an array,
    which is for instance important for the verification of GPU programs \cite{BlomH14}. \Iscs are complemented by unrestricted pure quantifiers over potentially heap-dependent expressions, which are
    essential for specifying functional properties.

%We also require that the expression $e(x)$ mentions the
%quantified variable exactly once (thus, $e(x)$ specifies some function
%from $x$ to receiver objects).

In this section, we present the first key ingredient of our symbolic execution technique: a representation for \iscs as part of
the verifier's symbolic state along with algorithms to manipulate
this representation.

%%--------------------------------------------------------------------
\subsection{Symbolic Heap Representation}\label{sec:heap_representation}
%%--------------------------------------------------------------------

As explained in \secref{smallfoot}, Smallfoot-style heap chunks
$\pointsto{o.f}[v,p]$ consist of a receiver value $o$, a field name $f$,
a location value $v$ and
a permission amount $p$. A \naive{} generalisation of this
representation would be to make $o$, $v$, and $p$ functions of the
bound variable of an \isc. However, such a representation
has severe drawbacks. Checking whether a heap chunk
provides permission to a location $y.f$ (challenge~2 above)
amounts to the existential query $\exists x. o(x) = y$;
SMT solvers provide poor support for such existential queries.
In the presence of fractional permissions, determining
\emph{how much} permission such a heap chunk provides
is worse still, requiring to calculate the sum of \emph{all}
$p(x_i)$ such that $x_i$ satisfies the existential query.

Our design avoids these difficulties with a simple restriction: we
require the receiver expressions $e(x)$ in an \isc{} to be
\emph{injective} in $x$, for all values of $x$ to which the \isc{}
provides permission. Under this restriction, we can soundly
assume that the mapping between the bound variable $x$ and
receiver expression $e(x)$ is \emph{invertible}
for such values, by some function $\inve$. We can then represent
an \isc{} over receivers $r = e(x)$ directly, essentially by
replacing $x$ by $\inve(r)$ throughout.

Our resulting design is to use \emph{quantified chunks} of
the form \chunk{r}{f}{\fvf}{p(r)}, in which $r$ (which is implicitly
bound in such a chunk) plays the role of a
quantified (reference-typed) receiver. Such a quantified chunk
represents $p(r)$ permission to all locations $r.f$; $p(r)$ may be any expression denoting a permission amount. The
\emph{domain} of a quantified chunk is the set of field locations
$r'.f$ for which $p(r') > 0$. The \emph{values} of these locations are
modelled by the function $\fvf$,
% \asout{ (from references to the type of
% field $f$)}
which we call a \emph{value map} and explain
in \secref{symbolic_values}. A \emph{symbolic heap} is a set of quantified chunks; a \emph{symbolic state} is a symbolic heap plus a set of path conditions, as usual.

\begin{figure}[t!]
\begin{myalgo}[H]
  \Inhale{$\hp_0$, $\pc_0$, \qpforall{$x\colon T$}{$c(x)$}{acc($e(x).f$, $p(x)$)}} $\compilesto$ \; \Indp
    Let $y$ be a fresh symbolic constant of type $T$ \;

    \tcc{Symbolically evaluate source-level expressions}
    \Var $(\pc_1, \sv{c}(y)) \gets \Eval{$\hp_0$, $\pc_0$, $c(y)$}$ \;
    \Var $(\pc_2, \sv{e}(y)) \gets \Eval{$\hp_0$, $\pc_1 \cup \{\sv{c}(y)\}$, $e(y)$}$ \;
    \Var $(\pc_3, \sv{p}(y)) \gets \Eval{$\hp_0$, $\pc_2$, $p(y)$}$ \;
    \Var $\pc_4 \gets \pc_3\, \setminus\, \{\sv{c}(y)\}$ \;
    \BlankLine
    \tcc{Introduce inverse function}
    Let $\svinve$ be a fresh function of type $T \rightarrow \reftype$ \;
    \Var $\pc_5\gets \pc_4 \cup \{\forall r\colon \reftype \cdot\ \sv{c}(\svinve(r)) \Rightarrow \sv{e}(\svinve(r)) = r\}$ \tcc*{\normalfont\textsc{(Inv-1)}}
    \Var $\pc_6 \gets \pc_5 \cup \{\forall x\colon T \cdot \sv{c}(x) \Rightarrow \svinve(\sv{e}(x)) = x\}$ \tcc*{\normalfont\textsc{(Inv-2)}}
    \BlankLine
    Let $\fvf$ be a fresh value map \;
    \Var $\hp_1 \gets \hp_0 \cup \{\chunk{r}{f}{\fvf}{\ite{\sv{c}(\svinve(r))}{\sv{p}(\svinve(r))}{\snone}}\}$ \;
    \Return $(\hp_1, \pc_6)$ \;
\Indm\BlankLine\BlankLine\BlankLine
  \Exhale{$\hp_0$, $\pc_0$, \qpforall{$x\colon T$}{$c(x)$}{acc($e(x).f$, $p(x)$)}} $\compilesto$ \; \Indp
    Let $y$ be a fresh symbolic constant of type $T$ \;
    \tcc{Symbolically evaluate source-level expressions}
    \Var $(\pc_1, \sv{c}(y)) \gets \Eval{$\hp_0$, $\pc_0$, $c(y)$}$ \;
    \Var $(\pc_2, \sv{e}(y)) \gets \Eval{$\hp_0$, $\pc_1 \cup \{\sv{c}(y)\}$, $e(y)$}$ \;
    \Var $(\pc_3, \sv{p}(y)) \gets \Eval{$\hp_0$, $\pc_2$, $p(y)$}$ \;
    \Var $\pc_4 \gets \pc_3\, \setminus\, \{\sv{c}(y)\}$ \;
    \BlankLine
    \tcc{Check injectivity of receiver expression}
    Let $y_1, y_2$ be fresh symbolic constants of type $T$ \;
    \Check $\pc_4 \vDash \sv{c}(y_1) \wedge \sv{c}(y_2) \wedge \sv{e}(y_1) = \sv{e}(y_2)\Rightarrow y_1 = y_2$ \; %\tcc*{Assert injectivity}
    \BlankLine
    \tcc{Introduce inverse function}
     Let $\svinve$ be a fresh inverse function of type $T \rightarrow \reftype$ \;
     \Var $\pc_5 \gets \pc_4 \cup \{\forall r\colon\reftype \cdot\ \sv{c}(\svinve(r)) \Rightarrow \sv{e}(\svinve(r)) = r\}$ \tcc*{\normalfont\textsc{(Inv-1)}}
     \Var $\pc_6 \gets \pc_5 \cup \{\forall x\colon T \cdot \sv{c}(x) \Rightarrow \svinve(\sv{e}(x)) = x\}$ \tcc*{\normalfont\textsc{(Inv-2)}}
    \BlankLine
    \tcc{Remove permissions}
    \Var $\hp_1 \gets \funSplitQP{$\hp_0$, $\pc_6$, $f$,
$(\lambda r \cdot \ite{\sv{c}(\svinve(r))}{\sv{p}(\svinve(r))}{\snone})$}$ \;
    % \quad $\funSplitQP{$\hp_0$, $\pc_2$, $\lambda r \cdot \inve(r)$, $\lambda x \cdot c(x)$, $\lambda x \cdot e(x)$, $\lambda x \cdot p(x)$, $f$}$ \; %% FULL VERSION
    \Return $(\hp_1, \pc_6)$
\end{myalgo}
\vspace{-1.5em}
\caption{Symbolic execution rules for inhaling and exhaling \iscs.
The \protect\Check{} instruction submits a query to the SMT solver. If the proof
obligation does not hold, it aborts
with a verification failure.
The \protect\Eval function evaluates an expression in a symbolic state and
yields updated path conditions and the resulting symbolic expression, see \secref{symbolic_values}. %PM: I think we should keep the reference because it is not clear why the path condition gets updated; since we probably do not want to explain this here, we should at least have a reference.
In both rules, the constraint $\sv{c}(y)$ is temporarily added to
the path conditions used
during the evaluation of $e(y)$ and $p(y)$; these expressions may
be well-formed only under this additional constraint.
}
\label{fig:inhale-exhale}
\end{figure}
\afterpage{\clearpage}

Under our injectivity restriction, we represent a source-level
assertion of the form \qpforall{$x\colon T$}{$c(x)$}{acc($e(x).f$,$p(x)$)} using a
quantified chunk of the form
\chunk{r}{f}{\fvf}{(\sv{c}(\svinve(r))\ ?\ \sv{p}(\svinve(r)) : 0)}
for a suitable value map $\fvf$ and inverse function $\svinve$.
Whenever necessary to avoid ambiguity, we use underlined expressions to denote the results of
symbolically evaluating corresponding source-level expressions;
with the exception of heap-dependent
expressions (see \secref{eval}), this evaluation is orthogonal
to the contributions of this paper.

Our injectivity restriction does not limit the data structures that
can be handled by our technique, provided specifications are
expressed appropriately. The restriction applies to
memory \emph{locations}, not to the \emph{values} stored in the
locations. Many examples such as \iscs ranging over
array indices or elements of a set naturally satisfy the
restriction. Ranges that may contain duplicates (for instance,
the fields of all objects stored in an array) can be
encoded by mapping them to a set
(thereby ignoring multiplicities) or by
using complex permission expressions $p$ that reflect
multiplicities appropriately.

%--------------------------------------------------------------------
\subsection{Inhaling and Exhaling Permissions}
\label{sec:inhale_exhale_permissions}
%--------------------------------------------------------------------

Using the symbolic heap design explained above, we define the
operations for inhaling and exhaling \iscs{} in
\figref{inhale-exhale}. The $\Inhale$
operation takes a symbolic heap $\hp_0$, path conditions $\pc_0$, and an \isc, and
returns an updated heap and path conditions.  Following the encoding
described in the previous subsection, the operation introduces a
(fresh) inverse function $\svinve$, which is constrained as the partial
inverse of the (evaluated) receiver expression $\sv{e}(x)$ by adding the constraints
\textsc{Inv-1} and \textsc{Inv-2} to the path conditions.
We will discuss controlling the instantiation of these quantifiers (and others introduced by our technique) in \secref{quantifiers}.
The fresh value map $\fvf$ models the (thus far
unknown) values of the heap locations in the domain of the new quantified chunk,
which is added to the symbolic heap $\hp_0$.

To encode our example (\figref{motivation}) in a tool without native array support, we model the array slots as a set of ghost objects, each with a field \inlinesilver{val} (representing the slot's value). That is, an array location $a[i]$ is modelled by the location
\inlinesilver{A(i).val}, where \inlinesilver{A} is an injective function mapping indices to these ghost objects.
Full details of the encoding of the running example are given 
online \cite[Example \inlinesilver{Parallel Array Replace}]{ViperOnline},
or in \appref{appendix_replace}.
Following \figref{inhale-exhale},
inhaling the second precondition (at the start of checking the method body) entails  introducing an inverse function $\inv{a}$ mapping array locations back to corresponding indices, and then adding a quantified chunk \chunk{r}{\sil{val}}{\fvf}{(\ite{\sv{{\sil{left}}} \leq \inva(r) <
    \sv{\sil{right}}}{\swrite}{\snone})}.
Correspondingly, at the program point after the two recursive calls, the symbolic heap will contain two quantified chunks: one for each array segment.

The $\Exhale$ operation is initially similar to $\Inhale$, one
difference being that the injectivity of the receiver expression is
checked before defining the inverse function. Removing
permissions is more complex than adding permissions because it may involve updates to
many existing quantified chunks in the symbolic state. This operation
is delegated to the auxiliary operation \funSplitQP, shown in
  \figref{split_qp}.

The injectivity check performed by \Exhale guarantees that the
  introduced inverse functions exist and satisfy the constraints added
  to the path conditions, which is required for soundness.
We assume here that each \Inhale operation has a corresponding 
\Exhale; for instance, inhaling a method precondition at the
beginning of a method body corresponds to exhaling the precondition
at the call site. Therefore, the check performed by \Exhale also 
covers the inverse functions introduced in corresponding \Inhale operations.

\begin{figure}[t]
\begin{myalgo}[H]
  \Fn{\funSplitQP{$\hp_0$, $\pc_0$, $f$, $q$}}{
    Let $\hp_f \subseteq \hp_0$ be all chunks in the given state for field $f$\;
    % \BlankLine
    \Var $\hp_f' \gets \emptyset$ \tcc*{Processed chunks}
    \Var $\ptt \gets q$ \tcc*{Permissions still to take}
    % \BlankLine
    \For{$(\chunk{r}{f}{\fvf_i}{q_i(r)}) \in \hp_f$}{
      \tcc{Determine the permissions to take from this chunk}
      \Var $\pt := (\lambda r \cdot \permmin(q_i(r), \ptt(r)))$\;
      \BlankLine\BlankLine
      \tcc{Decrease the permissions still needed}
      $\ptt := (\lambda r \cdot \ptt(r) - \pt(r))$\;
      \BlankLine\BlankLine
      \tcc{Add an updated chunk to the processed chunks}
      $\hp_f' := \hp_f' \cup \{\chunk{r}{f}{\fvf_i}{(q_i(r) - \pt(r))}\}$\;
    }
    \BlankLine\BlankLine
    \tcc{Check that sufficient permissions were removed}
    \Check $\pc_0 \vDash \forall r \cdot \ptt(r) = \snone$\;
    \BlankLine\BlankLine
%    \tcc{Field value function for the chunk to split off, and its definitional constraintss}
    % \hlLine{\Var $(\fvf, \eqs) \gets \funValueQP{$\hp_0$, $\lambda x \cdot c(x)$, $\lambda x \cdot e(x)$, $f$}$} %% FULL VERSION
%    \hlLine{\Var $(\fvf, \eqs) \gets \funValueQP{$\hp_0$, $f$}$}
%    \BlankLine
%    \tcc{Split-off chunk}
%    \Var $\symb{ch} \gets (\qch{\hl{\fvf(r)}}{q(r)})$\;
%    \BlankLine\BlankLine
%    \tcc{Return the updated symbolic heap}%, the chunk that has been split off, and the constraints for its \FVF}
    \Return $(\hp_0 \setminus \hp_f) \cup \hp'_f$%, \symb{ch}, \hl{\eqs})$
  }
\end{myalgo}
\vspace{-1.5em}
\caption{The remove operation. The argument $q$ maps references to permission
amounts. The operation checks that the symbolic heap contains at least $q(r)$
permission for each location $r.f$ and removes it.
%\vspace{-5mm}
}
\label{fig:split_qp}
\end{figure}

\funSplitQP takes as inputs an initial symbolic heap
$\hp_0$ and path conditions $\pc_0$, a field name $f$, and a function
$q$ that yields for each reference $r$ the permission amount for
location $r.f$ to be removed.  \funSplitQP fails with a verification
error if the initial heap does not contain the permissions in $q$, and
otherwise returns an updated symbolic state. This is achieved by
iterating over all available chunks for field $f$, greedily taking as
much of the still-required permissions ($\ptt$) as possible from the
current chunk ($\pt$).  Updating the chunks is expressed via
pointwise-defined functions describing the corresponding
permission amounts; they involve permission arithmetic, but no
  existential quantifiers, and can be handled efficiently by the underlying SMT
  solver.  After this iteration, \funSplitQP checks that all
requested permissions have been removed.

In our array example (\figref{motivation}), we exhale the second precondition
before each recursive call; this requires finding the appropriate permissions
from the (single) quantified chunk in the state at this point, and removing
them. Dually, when exhaling the postcondition at the end of the method body,
all permissions from both of the two quantified chunks yielded by the
recursive calls must be removed: the iteration in the \funSplitQP algorithm
achieves this.

Note that \funSplitQP's permission accounting is precise, which is
important for soundness and completeness: it maintains the invariant that (for
all $r$), the difference between the permissions held in the original state
and those requested via parameter $q$ is equal to the difference between those
held in the updated state and those still needed. If the operation succeeds,
we know (from the last check) that those still needed are exactly $0$, from
which we conclude that precisely the correct amounts were subtracted.

%--------------------------------------------
\subsection{Integrating Predicates with Iterated Separating Conjunctions}
\label{sec:iscs_and_predicates}
%--------------------------------------------
Predicates are a standard feature of verification tools for permission logics (including the Viper infrastructure on which our implementation is built); they integrate simply with our support for \iscs{}.
\figref{graph_predicate} shows an example of a predicate definition, parameterised by a set of nodes,
that defines a graph in terms of \iscs and closure properties over the
given set of nodes. Viper requires explicit ghost operations to
exchange a predicate instance $P(e)$ for its body (via an operation \inlinesilver`unfold $P(e)$`),
and vice versa
(via an operation \inlinesilver`fold $P(e)$`);
this is a standard way to handle possibly-recursive predicates.
In terms of the underlying verifier, an operation \inlinesilver`fold $P(e)$`
essentially corresponds to \inlinesilver`exhale $P_{body}(e)$` followed by
\inlinesilver`inhale $P(e)$`, and dually for \inlinesilver`unfold $P(e)$`.
Since our support for \iscs is expressed in terms of inhale and exhale
rules, it naturally integrates with Viper's existing way of handling
predicates; our implementation supports predicates with \iscs{} 
and pure quantifiers in their bodies,
as illustrated by the \code{graph} predicate.

Our implementation does not yet support predicates inside \iscs,
but our presented technique extends straightforwardly to support this.
Inhaling an \isc which ranges over predicate instances yields, just as for accessibility predicates for fields, a new quantified chunk. An \inlinesilver`unfold` of a predicate belonging to such a chunk can be handled
by
exhaling the predicate instance (removing it from the chunk's permissions), and then inhaling the predicate's body. Folding an instance 
inhales a quantified predicate chunk that provides permissions to the
single instance. We plan to extend our implementation to also support this feature combination, which will allow one to denote an unbounded 
number of predicate instances.

\begin{figure}[t]
\begin{silver}[mathescape=true]
predicate Graph(nodes: Set[Ref]) {
     (forall n: Ref :: n in nodes ==> acc(n.left))
  && (forall n: Ref :: n in nodes ==> acc(n.right))
  && (forall n: Ref :: n in nodes && n.left != null ==> n.left  in nodes)
  && (forall n: Ref :: n in nodes && n.right != null ==> n.right in nodes)
}
\end{silver}
\vspace{-0.75em}
\caption{A predicate defining a graph in terms of \iscs and closure properties over a given set of nodes (that form the graph).% The \inlinesilver{Set} type is already supported by the Viper tools.
}
%\vspace{-1.0em}
\label{fig:graph_predicate}
\end{figure}

%\ms{Our implementation already supports predicates that encapsulate \iscs and
%pure quantifiers in their bodies, as illustrated by the \code{graph}
%predicate. With the extension described above, it will also be possible to
%specify unboundedly many instances of such predicates.}

%%%%%%%%%%%%%%%%%%%%%%%%%%%%%%%%%%%%%%%%%%%%%%%%%%%%%%%%%%%%%%%%%%%%%
\section{Treatment of Symbolic Values}
\label{sec:symbolic_values}
%%%%%%%%%%%%%%%%%%%%%%%%%%%%%%%%%%%%%%%%%%%%%%%%%%%%%%%%%%%%%%%%%%%%%

So far we have addressed the first three technical challenges
  described in \secref{background} by presenting a novel heap
  representation for \iscs together with algorithms that let the
  verifier efficiently add, as well as check for and remove
  permissions. In this section we present our solution to the remaining two
challenges, concerned with the evaluation and framing of expressions.

%--------------------------------------------------------------------
\subsection{Symbolic Evaluation of Heap-Dependent Expressions}
\label{sec:eval}
%--------------------------------------------------------------------
Quantified chunks \chunk{r}{f}{\fvf}{q(r)} represent value information
via the value map $\fvf$. The existence of such a chunk in a symbolic
heap allows the evaluation of a read of field $f$ for any receiver
in the domain of the heap chunk, to an application of the value
map. Intuitively, $\fvf$ represents a partial function from this
domain to values (of the type of the field $f$). Since SMT solvers
typically do not natively support partial functions, we model value
maps as under-specified total functions from the receiver reference
(the field $f$ is fixed) to the type of $f$. We
\emph{apply} these functions only to references whose $f$ field location is
in the chunk's domain.  This is why the
 $\Exhale$ algorithm (\figref{inhale-exhale}) does not need to explicitly remove
information about the values stored in the locations whose permissions
are removed;
the underlying total function still represents appropriate values
for the new
(smaller) domain.

\subsubsection{Summarising Value Maps.}
%--------------------------------------------------------------------
Inhaling permissions adds a fresh heap chunk with a fresh value map
(see \figref{inhale-exhale}). Therefore, a symbolic heap may contain multiple chunks for the same field, each with its own value map. In the
presence of fractional permissions, the domains of these chunks may
overlap such that the value of one location $x.f$ may be represented by
multiple value maps. Similarly, the value of $x.f$ may be represented
by multiple maps when the receiver $x$ is quantified over and the
permissions to different instantiations of the quantifier are recorded
in different chunks. Therefore, all of these value maps need to be
considered when evaluating such a field access.

\begin{figure}[t]
\begin{myalgo}[H]
  \Fn{\funValueQP{$\hp_0$, $f$}}{
    Let $\hp_f \subseteq \hp_0$ be all quantified chunks in the given heap for field $f$\;
    Let $\fvf$ be a fresh \FVF \;
    \Var $\eqs \gets \emptyset$ \tcc*{Value summary path conditions}
    \Var $\perms \gets \lambda r \cdot 0$ \tcc*{Permission summary}
    \For{$(\chunk{r}{f}{\fvf_i}{q_i(r)}) \in \hp_f$}{
      $\eqs \gets \eqs \cup \{\forall r \cdot \snone < q_i(r) \Rightarrow \fvf(r) = \fvf_i(r)\}$ \tcc*{\normalfont(\VmDefEq)}
      $\perms \gets \lambda r \cdot (\perms(r) + q_i(r))$
    }
    \Return $(\fvf, \eqs, \perms)$
  }
\end{myalgo}
\vspace{-1.5em}
\caption{The $\protect\funValueQP$ operation introduces a fresh \FVF for field $f$ and
constrains it according to the \FVFs of all heap chunks for $f$. It also returns a function summarising the permissions held for the field $f$.}
%\vspace{-1.0em}
\label{fig:value_qp}
\end{figure}

In order to incorporate information from all relevant chunks,
and provide a simple translation for field-lookups,
we summarise the value maps for all chunks for a field $f$ \emph{lazily} before
we evaluate an expression $e.f$.
This summarisation is defined by the $\funValueQP$ operation in
\figref{value_qp}.  For each quantified chunk with the appropriate
field, it equates a newly-introduced value map with the value map in
the chunk at all locations in the chunk's domain. Analogously, it
builds up a permission expression summarising the permissions held per receiver, across all
heap chunks for the field $f$; we use this permission expression to
check whether a field access is permitted.  

Note that the definition
of $\funValueQP$ does not depend on path conditions, only on the
symbolic heap; it can be computed without querying the SMT solver.
Our implementation memoizes $\funValueQP$, avoiding the
duplication of the function declarations and path conditions
defining the value and permission maps.

\subsubsection{Symbolic Evaluation.}
%--------------------------------------------------------------------
Symbolic evaluation of expressions is defined by an operation \Eval,
which takes a symbolic heap, path conditions, and an expression, and
yields updated path conditions and the symbolic value of the
expression; the cases for field lookup and pure quantifiers are given
in \figref{rules_eval} (some additional cases can be found in
\appref{rules}).  Using the $\funValueQP$ operation, we can simply
define the evaluation of a field lookup, as shown first in
\figref{rules_eval}. To evaluate such an expression, we check that at
least some permission to the field location is held in the current
symbolic heap, and use the value map generated by $\funValueQP$ to
define the value of the field lookup. Via the path conditions
generated by $\funValueQP$, any properties known about the value maps
of any of the corresponding quantified chunks will also be
known about the resulting symbolic value.  In each reachable
  state, these properties are consistent, which implies in particular
  that there exists a value for the field lookup that satisfies all of
  them. Viper regularly checks for inconsistent path conditions
  and prunes the current program path if it detects an unreachable
  state.

% removed from check line: \funPermQP{$\hp_0$, $f$}$

Evaluating pure quantifiers is handled by replacing the bound variable
with a fresh constant and evaluating the quantifier body. Additional
path conditions generated during this recursive evaluation might
mention the fresh constant; these are universally
quantified over when returning the path conditions.
%Note
%that we could also universally quantify over each such path condition
%individually; we avoid this choice, since it would result (in general)
%in more quantifier instantiations for the SMT solver.

\subsubsection{Inhale, Exhale, and Field Writes.}
%--------------------------------------------------------------------
Inhaling and exhaling pure boolean expressions is implemented by first
symbolically evaluating the expression
and then either adding the resulting symbolic expression to the path
conditions or checking it, respectively (see \appref{rules}).

A field write
\inlinesilver{$e_1$.$f$ := $e_2$} is desugared
as: \inlinesilver{exhale acc($e_1$.$f$);} \inlinesilver{inhale acc($e_1$.$f$);}
\inlinesilver{inhale $e_1.f$ == $e_2$}. The exhale checks that the heap has the required
permission and removes it; the inhales create a new chunk with the
previously-removed permission and constrain the associated \FVF such
that it maps receiver $\sv{e_1}$ to the value of $\sv{e_2}$.
For example, the field write \inlinesilver{a[left] := to} in
\figref{motivation} is executed in a symbolic heap with a single \ic
that provides full permissions to each array location. After the field
write has been executed, the heap contains two \ics: the initial one,
still providing full permissions to each array location \emph{except for}
\inlinesilver{a[left]} (and with an unchanged value map), and a second one that provides permissions to
\inlinesilver{a[left]} only, with a fresh value map
representing the updated value.

\begin{figure}[t]
\begin{myalgo}[H]
  \Eval{$\hp_0$, $\pc_0$, $e.f$} $\compilesto$ \; \Indp
    \Var $(\pc_1, \sv{e}) \gets \Eval{$\hp_0$, $\pc_0$, $e$}$ \;
    \Var $(\fvf, \eqs, \perms) \gets \funValueQP{$\hp_0$, $f$}$ \;
    \Check $\pc_1 \vDash 0 < \perms(\sv{e})$ \;
    \Return $(\pc_1 \cup \eqs, \fvf(\sv{e}))$ \;
\Indm\BlankLine\BlankLine\BlankLine
  \Eval{$\hp_0$, $\pc_0$, \textup{\sforall} $x$ :: $e(x)$} $\compilesto$ \; \Indp
    Let $y$ be a fresh symbolic constant \;
    \Var $(\pc_1,\sv{e}(y)) \gets \Eval{$\hp_0$, $\pc_0$, $e(y)$}$ \;
    \Return $(\{b\in\pc_1 \mid y\notin \FV{b}\} \cup \{ \forall x\cdot(\bigwedge_{b\in\pc_1, y\in\FV{b}} \sub{b}{x}{y}) \},\, \forall\, x \cdot \sv{e}(x))$ \;
%% \Indm\BlankLine\BlankLine\BlankLine
%%   \Eval{$\hp_0$, $\pc_0$, $e_1 \otimes e_2$} $\compilesto$ \; \Indp
%%     \Var $(\pc_1,\sv{e_1}) \gets \Eval{$\hp_0$, $\pc_1$, $e_1$}$ \;
%%     \Var $(\pc_2, \sv{e_2}) \gets \Eval{$\hp_0$, $\pc_2$, $e_2$}$ \;
%%     \Return $(\pc_2, \sv{e_1} \otimes \sv{e_2})$ \;
\end{myalgo}
\vspace{-1.5em}
\caption{Symbolic evaluation of field reads and pure quantifiers.}
\label{fig:rules_eval}
\end{figure}

%%--------------------------------------------------------------------
\subsection{Framing Heap-Dependent Expressions}
\label{sec:abstractions}
%%--------------------------------------------------------------------

Permissions provide a straightforward story for framing the values of heap locations (and pure quantifiers over these): so
long as the symbolic state contains \emph{some} permission to a field location, its value will be preserved.
However, framing heap-dependent functions is more complicated
\cite{SmansJP10,HeuleKassiosMuellerSummers13}. The value of a function
can be framed so long as all locations the function depends on remain
unchanged. To express a function's dependency on the heap, its
precondition must require permission to all locations its
implementation may read. For any given function application, the symbolic
values of these locations are called the \emph{snapshots} of the
function application. Consequently, two function applications yield
the same result if they take the same arguments and have equal
snapshots. One can thus model a heap-dependent function at the SMT
level by a function taking snapshots as additional arguments \cite{SmansJP10}.

\Iscs complicate this approach because a function whose precondition
contains an \isc may depend on an unbounded set of heap locations.
The values of these locations cannot be represented by a fixed number
of snapshots. It is also not possible to represent them as a value map
since these are modelled at the SMT
level as total functions, causing two problems.  First, requiring
  equality of total functions would include locations the heap-dependent function does not actually depend on; since the values for these locations are under-specified, the
equality check would often fail even when the function value could be
soundly framed.
Second, a function cannot be used as a function argument, nor
compared for equality in the first-order logic supported by SMT solvers.

We address the first problem by modelling snapshots as
\emph{partial} functions called \emph{partial value maps}, and the
second by applying \emph{defunctionalisation} \cite{Reynolds1972}. That is, we model a partial value map for a
field $f$ of type $T$ as a value of an (uninterpreted) type
$\fvftype$, together with a function $\domain \colon \fvftype
\rightarrow \settype{\reftype}$ for the domain of the partial value
map, and a function $\lookup\colon \fvftype \times \reftype
\rightarrow T$ for the result of applying a partial value map to a
receiver reference.  We also include an extensionality
  axiom for partial value maps, allowing us to prove equality when
two partial value maps are equal as partial functions.

Following the prior work, we model a heap-dependent function via a
function at the SMT level, with a partial value map as additional
snapshot argument for each \isc{} required in the function's
precondition.  For each
application of such a function, we check that the current state
contains all permissions required by the function precondition.
If this is the case, we process each \isc in the precondition in
turn. For an \isc for a field $f$, we employ the
$\funValueQP$ operation (\figref{value_qp}) to summarise the value
information $\vs$ for the field $f$ in the current symbolic state, and
introduce a fresh constant $\pvs$ of type
$\fvftype$. We constrain $\domain(\pvs)$ to yield
the set of references in the domain of the \isc, and
for all receivers $r$ in this domain, assume $\lookup(\pvs,r) =
\vs(r)$. $\pvs$ is then
used as a snapshot argument to the translated function.
%Note that pure
%assertions required by the function's precondition can also be
%asserted to hold using the summarised value information.

%%%%%%%%%%%%%%%%%%%%%%%%%%%%%%%%%%%%%%%%%%%%%%%%%%%%%%%%%%%%%%%%%%%%%
\section{Controlling Quantifier Instantiations}
\label{sec:quantifiers}
%%%%%%%%%%%%%%%%%%%%%%%%%%%%%%%%%%%%%%%%%%%%%%%%%%%%%%%%%%%%%%%%%%%%%

When generating quantifiers for an SMT solver, it is important to
carefully control their instantiation
\cite{HeuleKassiosMuellerSummers13,LeinoM09,Moskal09} by
  providing syntactic triggers. A quantifier
$\forall x \cdot P(x)$ may be decorated with a \emph{trigger}
$\{f(x)\}$, which instructs the solver to
instantiate $x$ with a term $e$ only if $f(e)$ is a term encountered
by the solver during the current proof effort. Triggers must be
  chosen carefully: enabling too few instantiations may cause examples
  to fail unexpectedly, while too many may lead to unreliable
  performance or even non-termination of the solver
 (see also~\secref{implementation}).

\begin{figure}[t]
\begin{myalgo}[H]
  $\forall r\colon \reftype \cdot \,\,\,\triggers{\fvf(r)}\triggers{\fvf_i(r)}\,\,\, \ \snone < q_i(r) \Rightarrow \fvf(r) = \fvf_i(r)$ \tcc*{\normalfont(\VmDefEq)}

  $\forall r\colon \reftype \cdot \,\,\,\triggers{\svinve(r)}\,\,\, \ \sv{c}(\svinve(r)) \Rightarrow \sv{e}(\svinve(r)) = r$ \tcc*{\normalfont\textsc{(Inv-1)}}

    $\forall x\colon T \cdot \,\,\,\triggers{e(x)}\,\,\,\ \sv{c}(x) \Rightarrow \svinve(\sv{e}(x)) = x$ \tcc*{\normalfont\textsc{(Inv-2)}}
\end{myalgo}
\vspace{-1.5em}
\caption{Example triggers used in our SMT encoding.}
\label{fig:axioms_triggers}
\end{figure}

We carefully select triggers for all quantifiers generated by
our technique (although we have omitted them from the presentation so far).
\figref{axioms_triggers} shows three representative examples.
The path condition \VmDefEq relates the value map introduced by the
$\funValueQP$ operation to the value maps of heap chunks
(\figref{value_qp}).  The two triggers express
alternatives: they allow instantiating the path condition if \emph{either} of
the two value maps have been applied to the term
instantiating $r$. This design
%is important because of how our
%$\funValueQP$ operation (\figref{value_qp}) relates the
%newly-generated value map with each quantified chunk individually;
%this indirectly
allows us to derive relationships between two evaluations of an expression, which introduce two summary value maps.
Instantiating \VmDefEq in both directions allows us to relate
these value maps via the value maps of heap chunks.

The next two examples define the inverse function of a receiver
  expression (see \figref{inhale-exhale}). The trigger $\svinve(r)$
  for \textsc{Inv-1} is essential for relating occurrences of the
  inverse function to the original expression $e$.
%Recall
%  (\figref{split_qp}) that \iscs{} are translated into quantified
%  chunks in which $e$ does \emph{not} occur, while its inverse
%  function does.
%
The case of \textsc{Inv-2} is almost symmetrical, but with extra
technicalities. Since $e$ comes from the source program, it may not be
an expression allowed as a trigger. Trigger terms must typically
include at least one function application (if $e(x)$ were simply $x$,
this could not be used), and no built-in operators such as
addition. In the former case, we use $\fvf(x)$ as a trigger, where
$\fvf$ is the value map of the relevant chunk; the quantifier will then
be instantiated whenever we look up a value from the
chunk, which is when we need the definition of the inverse
  function. In the latter case, we resort to allowing the underlying
tools select trigger terms, which may lead to incompleteness.
However, we did not observe any such incompletenesses in our experiments.

Instantiating either of the two axioms \textsc{Inv-1} and \textsc{Inv-2} gives rise to potentially new function application terms suitable for triggering the other axiom. For example, when instantiating \textsc{Inv-2} due to a term of the shape $e(x)$, we learn the equality $\svinve(\sv{e}(x)) = x$ in which the function application $\svinve(\sv{e}(x))$ matches the trigger for the \textsc{Inv-1} axiom. Instantiating this axiom, in turn, will provide the equality $\sv{e}(\svinve(\sv{e}(x))) = \sv{e}(x)$. Note however, that this will not cause an indefinite sequence of instantiations of these two axioms (a so-called matching loop): SMT solvers consider quantifier instantiations \emph{modulo} known equalities. Thus, the function application $\sv{e}(\svinve(\sv{e}(x)))$ does not give rise to a \emph{new} instantiation of \textsc{Inv-2}, since the term to be matched against the quantified variable ($\svinve(\sv{e}(x))$) is already known to be equal to $x$, which was used for the prior instantiation.

%% ***********************************************************************
\section{Evaluation}
\label{sec:implementation}
%% ***********************************************************************

We have implemented our technique as an extension of the Viper
verification infrastructure \cite{MuellerSchwerhoffSummers16};
the implementation is open source and can be tried online~\cite{ViperOnline}.
%\asout{Viper is
  %open-source, and can be obtained from
 % \textbf{\url{http://viper.ethz.ch}}.}
%, together with several examples and lots of regression tests.
%
To evaluate the
performance of our technique, we ran experiments
  with three kinds of input programs: (1)~9
hand-coded verification problems
involving arrays and graphs, including our running example
(see the Viper examples page~\cite{ViperOnline} or \appref{appendix_descriptions} for details,
and \appref{appendix_examples} for two encodings), 
(2)~65
examples generated by the VerCors
project at the University of Twente
\cite{BlomH14}, which uses our implementation to
encode GPU verification problems,
% \asout{ involving arrays}
and (3)~82 additional regression tests.

\begin{figure}[t]
\setlength{\tabcolsep}{4pt}
\newcolumntype{d}[1]{D{.}{.}{#1}}
\newcommand{\h}[1]{\multicolumn{1}{c}{#1}}
\newcommand{\hb}[1]{\multicolumn{1}{c|}{#1}}
\begin{center}
\begin{tabular}{l|d{3}|d{2.2}|d{2.2}|d{2.2}}
  Program & \hb{Size (LOC)} & \hb{Time (s)} & \hb{w/o memoization} & \h{w/o triggers} \\
  \hline
  \texttt{arraylist}              & 114 & 1.93 & -7.29\% & -16.53\% \\
  \texttt{quickselect}            & 132 & 2.51 & +24.44\% & -4.23\% \\
  \texttt{binary-search}          & 47  & 0.31 & +14.15\% & -8.94\% \\
  \texttt{graph-copy}             & 120 & 1.81 & +14.93\% & +21.21\% \\
  \texttt{graph-marking}          & 53  & 1.71 & +41.29\% & -30.95\% \\
  \texttt{longest-common-prefix}  & 34  & 0.19 & +6.51\%  & -10.73\% \\
  \texttt{max-elimination}        & 59  & 0.50 & +45.41\% & -0.07\% \\
  \texttt{max-standard}           & 53  & 0.24 & +16.40\% & +2.43\% \\
  \texttt{parallel-replace}       & 56  & 0.27 & +3.71\%  & -6.12\% \\
\end{tabular}
\end{center}
\vspace{-1.0em}
\caption{Performance evaluation of our implementation on verification challenges.
Lines of code (LOC) does not include blank lines and comments.
Column ``Time (s)'' gives runtimes of the base version of our implementation;
columns ``w/o memoization'' and ``w/o triggers'' show the \% difference in time relative to the base version.
%\vspace{-1.0em}
}
\label{fig:runtimes_examples}
\end{figure}

\figref{runtimes_examples} shows the results for (1), and
\figref{runtimes_regressions} those for (2) and (3).
We performed our experiments
on an Intel Core i7-4770 3.40GHz
with 16GB RAM machine running Windows 7 x64 with an SSD\@.
The reported times are averaged over 10 runs of each verification
(with negligible standard deviations). Timings do not include
JVM start-up: we persist a JVM across test runs using the Nailgun tool.

\begin{figure}[t]
\setlength{\tabcolsep}{4pt}
\newcolumntype{d}[1]{D{.}{.}{#1}}
\newcommand*{\mc}[3]{\multicolumn{#1}{#2}{#3}}
\newcommand*{\h}[1]{\makecell{#1}}
\begin{center}
\begin{tabular}{l|c|c|cc|cc|cc}
  & No. & Size & \mc{2}{c|}{Time} & \mc{2}{c|}{w/o memoization} & \mc{2}{c}{w/o triggers} \\
  \h{\ \\Program Set } & \h{Files \\ (\#)} & \h{Mean \\ (LOC)} & \h{Mean \\ (s)} & \h{Max \\ (s)} & \h{Mean \\($\pm$) } & \h{Max \\ (s)} & \h{Mean \\($\pm$) } & \h{Max \\ (s)} \\
  \hline
  VerCors     & 65 & 104 & 0.72 & 11.81 & +0.92\% & 15.71 & -4.40\% & 8.83 \\
  Regressions & 82 & 34  & 0.22 & 3.41  & +0.58\% & 3.81  & -2.24\% & 3.38 \\
\end{tabular}
\end{center}
\vspace{-1.0em}
\caption{Performance evaluation of our implementation on two sets of
programs: the ``VerCors'' set contains (non-trivial) programs generated
by the VerCors tool, ``Regressions'' contains (usually simple) regression
tests; column ``No.~Files'' displays the number of files per program set.
All input files are available as part of the Viper test suite.
}
%\vspace{-1.0em}
\label{fig:runtimes_regressions}
\end{figure}

Our experiments show that our implementation
is consistently fast: all examples verify in a few seconds.
% (we also
%observed consistent runtimes, that is, negligible standard
%deviations).  
Since SMT encodings sometimes exhibit worse performance
for \emph{failed} verification attempts, we also tested 4 variants of
each example from \figref{runtimes_examples} in which we seeded
errors; in all cases the errors were detected with lower runtimes (the
verifier halts as soon as an error is detected).

To measure the effect of memoizing calls to $\funValueQP$, we disabled
this feature and measured the difference in runtimes over the same
inputs. As shown
in the ``w/o memoization'' columns,
disabling this optimisation typically increases the runtime,
but not enormously; a likely explanation for the relatively small
difference is that $\funValueQP$ performs
the iteration over quantified chunks efficiently, without querying
the SMT solver. The
number of quantified chunks in a given symbolic state is also
typically kept small: the tool performs modular verification per
method/loop body, and we eagerly remove any quantified chunks that no
longer provide permissions (after an exhale).

To evaluate the importance of our use of triggers for controlling
quantifier instantiations (see \secref{quantifiers}), we also compare
with a variant of our implementation in which triggers are omitted,
leaving this task to the underlying tools (that is, Viper and Z3~\cite{MouraB08}). The relative times are
shown in the ``w/o triggers'' columns. We observe that this variant
typically \emph{improves} verification time. However, the triggers chosen automatically by Viper and Z3 are too strict: 7\% of the
programs (11 out of the 156 original programs) fail
  spuriously in this version. This, as well as a general reduction in
  quantifier instantiations, explains the effect on the runtime: the
  longest-running example in our base implementation (averaging
  11.82s) takes only 3s without our triggers, but wrongly fails to
  verify.  The longest-running example in the variant without
  triggers takes 8.83s but also has a high standard deviation of
  4.71s, suggesting that performance also becomes unpredictable
when triggers are selected automatically.
The triggers that we choose thus avoid spurious errors
  and provide predictable, fast performance.

%%%%%%%%%%%%%%%%%%%%%%%%%%%%%%%%%%%%%%%%%%%%%%%%%%%%%%%%%%%%%%%%%%%%%
\section{Conclusions and Future Work}
\label{sec:conclusion}
%%%%%%%%%%%%%%%%%%%%%%%%%%%%%%%%%%%%%%%%%%%%%%%%%%%%%%%%%%%%%%%%%%%%%

We have presented the first symbolic execution technique that supports
\iscs{}. This feature provides the possibility of specifying
random-access data structures and provides an alternative mechanism to
recursive definitions which is essential in the common case when a
data structure can be traversed in multiple ways. Our technique
generalises Smallfoot-style symbolic execution and is, thus, applicable
to other verifiers for permission logics
using this common implementation technique.

Two of the authors participated in the recent VerifyThis verification
competition at ETAPS'16 (see \url{http://etaps2016.verifythis.org/})
using our implementation, and won the \emph{Distinguished User-assistance 
Tool Feature} for the \isc{} support described in this paper: this prize was awarded for a feature that proved particularly useful
during the competition.

As future work, we plan to build on our verification technique in four
ways. First, we plan to extend our technique to support predicates
under \iscs, as discussed in \secref{iscs_and_predicates}. Second,
we plan to combine our verification technique with
inference techniques that make use of \iscs, such as the shape
analysis developed by Lee \etal{}~\cite{LeeYY05}.
Third, we plan to support \inlinesilver{foreach} statements that
perform an operation (\eg, unfolding a predicate) on each instance
of a quantifier without requiring a loop (and invariant). Such statements
require permissions that can be expressed using \iscs. Fourth,
we plan to integrate support for aggregates in
pure assertions \cite{LeinoM09}, which provide
another means for specifying functional properties over locations
described by an \isc.

%Our implementation does not yet support predicates inside an \isc{}
%assertion, but our techniques extend directly to this case. Adapting
%analogous support to that for field locations (supporting fold/unfold
%of predicates similarly to reading/writing fields in the program) is
%straightforward, and we plan to implement this.

%
%The \funSplitQP algorithm could be useful
%
%is also useful in verifiers that do not
%support \isc. Even with just conventional heap chunks, exhaling permission
%to a single location may affect several chunks if they contain fractional
%permissions or when it is not statically known which chunk contains
%the permission because of disjunctive aliasing information. Existing
%verifiers solve the former problem via heap compression, a
%(potentially expensive)
%operation that tries to combine multiple chunks for the same location.
%For the latter, they require explicit case splits (expressed via ghost
%statements) to resolve the aliasing.
%

%\newpage

\bibliographystyle{abbrv}
\bibliography{quantified_permissions}

\begin{appendix}
%%%%%%%%%%%%%%%%%%%%%%%%%%%%%%%%%%%%%%%%%%%%%%%%%%%%%%%%%%%%%%%%%%%%%
\section{Additional Definitions and Symbolic Execution Rules}
\label{sec:rules}
%%%%%%%%%%%%%%%%%%%%%%%%%%%%%%%%%%%%%%%%%%%%%%%%%%%%%%%%%%%%%%%%%%%%%

\subsubsection{Partial Value Maps.}

\figref{preamble} shows background definitions related to partial value maps
(see \secref{abstractions}), which are emitted to the SMT solver before the
verification starts. The background definitions include a type
$\fvftype$ and, per field declaration, a function $\domain$ that denotes the domain
of a partial value map, a function $\lookup$ that denotes applying a partial
value map to a receiver to obtain the value of the corresponding field
location, and an extensionality axiom stating that two partial value maps are
equal if their domains agree and if they agree on the values in their domain.

\begin{figure}[t]
\newcommand*{\FD}{\symb{FD}\xspace}
\newcommand*{\fields}{\symb{fields}\xspace}
\newcommand*{\types}{\symb{types}\xspace}
\begin{myalgo}[H]
  1. Let \FD be the set of all field declarations $f\colon T$ of a given program for which \iscs are used \;
  \BlankLine
  2. Declare a type $\fvftype$ \;
  \BlankLine
  3. Declare a function $\domain\colon \fvftype \rightarrow \settype{\reftype}$ per declaration $f\colon T \in FD$ \;
  \BlankLine
  4. Declare a function $\lookup\colon \fvftype \times \reftype \rightarrow T$ per declaration $f\colon T \in FD$ \;
  \BlankLine
  5. Declare the following extensionality axiom per declaration $f\colon T \in FD$: \; \Indp
    $\forall \pvs_1, \pvs_2\colon \fvftype \cdot$ \triggers{\toSnap(\pvs_1),\toSnap(\pvs_2)} \; \Indp
      $\ \ \domain(\pvs_1) = \domain(\pvs_2) \wedge$ \;
      $\forall r\colon \reftype \cdot r \in \domain(\pvs_1) \Rightarrow \lookup(\pvs_1, r) = \lookup(\pvs_2, r)$ \;
      $\Rightarrow \pvs_1 = \pvs_2$ \;
  \Indm \Indm
\end{myalgo}
\vspace{-1.5em}
\caption{Background definitions related to partial value maps
(see \secref{abstractions}). $\domain$ denotes the domain of a partial value
map, $\lookup$ its application to a reference.}
\label{fig:preamble}
\end{figure}

The trigger of the extensionality axiom
$\triggers{\toSnap(\pvs_1),\toSnap(\pvs_2)}$ ensures that the
extensionality axiom is instantiated whenever it is necessary to
reason about the equality of partial value maps that are used as
snapshots. Wrapping partial value maps by $\toSnap$ is necessary
because Viper requires snapshots to uniformly be of type $\snaptype$;
function $\toSnap$ embeds values into the $\snaptype$ type (a
corresponding inverse function exists as well).  This external
requirement (of Viper, not of our technique) turned out to be
beneficial for us, since it allows choosing triggers that are
permissive, yet yield good performance.

\subsubsection{Inhaling and Exhaling Pure Assertions.}

\figref{inhale_exhale_pure} shows the symbolic execution rules for inhaling
and exhaling potentially heap-dependent (but pure) assertions
such as pure quantifiers. Both rules use \inlinesilver{eval} to
evaluate the assertion; the result is then added to the path
conditions or asserted to hold in the current state, respectively.

\begin{figure}[t]
\begin{myalgo}[H]
  \Inhale{$\hp_0$, $\pc_0$, $e$} $\compilesto$ \; \Indp
    \Var $(\pc_1, \sv{e}) \gets \Eval{$\hp_0$, $\pc_0$, $e$}$ \;
    \Return ($\hp_0, \pc_1 \cup \{\sv{e}\}$) \;
\Indm\BlankLine\BlankLine\BlankLine
  \Exhale{$\hp_0$, $\pc_0$, $e$} $\compilesto$ \; \Indp
    \Var $(\pc_1, \sv{e}) \gets \Eval{$\hp_0$, $\pc_0$, $e$}$ \;
    \Check $\pc_1 \vDash \sv{e}$ \;
    \Return ($\hp_0, \pc_1$) \;
\end{myalgo}
\vspace{-1.5em}
\caption{Symbolic execution rules for inhaling and exhaling pure
assertions.}
\label{fig:inhale_exhale_pure}
\end{figure}

\subsubsection{Symbolic Evaluation of Expressions.}

\figref{rules_eval_more} shows selected symbolic execution rules for
evaluating expressions. Evaluating an implication $e_1 \Rightarrow e_2$
starts by evaluating $e_1$, and temporarily assuming $\sv{e_1}$ while
evaluating $e_2$ (see also the discussion of
\figref{inhale-exhale} in \secref{heap_representation}).
From the path conditions obtained from evaluating $e_1$ ($\pi_{\delta}$),
all instances of \VmDefEq are extracted ($\pi_v$).
The final set of path conditions, with which the
verification proceeds ($\pi_3$), includes the path conditions obtained
from the evaluation of $e_1$, all instances of \VmDefEq that were obtained
from evaluating $e_2$ (this allows memoizing \inlinesilver{summarise} because
value map definitions are always in scope, that is, are not nested under implications),
and --- conditionally on $\sv{e_1}$ --- the remaining path conditions from
evaluating $e_2$.

\begin{figure}[t]
\begin{myalgo}[H]
  \Eval{$\hp_0$, $\pc_0$, $e_1 \Rightarrow e_2$} $\compilesto$ \; \Indp
    \Var $(\pc_1, \sv{e_1}) \gets \Eval{$\hp_0$, $\pc_0$, $e_1$}$ \;
    \Var $(\pc_2, \sv{e_2}) \gets \Eval{$\hp_0$, $\pc_1 \cup \{e_1\}$, $e_2$}$ \;
    \Var $\pc_{\delta} \gets \pc_2 \setminus (\pc_1 \cup \{e_1\})$ \;
    \Var $\pc_{v} \gets \{b \in \pc_{\delta} \mid b \text{ is instance of \VmDefEq} \}$ \;
    \Var $\pc_3 \gets \pc_1 \cup \pc_{v} \cup \{\sv{e_1} \Rightarrow \bigwedge (\pc_{\delta} \setminus \pc_{v}) \}$ \;
    \Return $(\pc_3, \sv{e_1} \Rightarrow \sv{e_2})$ \;
\Indm\BlankLine\BlankLine\BlankLine
  \Eval{$\hp_0$, $\pc_0$, $\fun(e_1, \ldots, e_n)$} $\compilesto$ \tcc*{$\fun$ is heap-independent} \Indp
    \Var $(\pc_1,\sv{e_1}) \gets \Eval{$\hp_0$, $\pc_0$, $e_1$}$ \;
    $\ldots$ \;
    \Var $(\pc_n,\sv{e_n}) \gets \Eval{$\hp_0$, $\pc_{n-1}$, $e_n$}$ \;
    \Return $(\pc_n, \sv{\fun}(\sv{e_1}, \ldots, \sv{e_n}))$ \;
\Indm\BlankLine\BlankLine\BlankLine
  \Eval{$\hp_0$, $\pc_0$, $e_1 \wedge e_2$} $\compilesto$ \; \Indp
    \Var $(\pc_1,\sv{e_1}) \gets \Eval{$\hp_0$, $\pc_0$, $e_1$}$ \;
    \Var $(\pc_2, \sv{e_{\Rightarrow}}) \gets \Eval{$\hp_0$, $\pc_1$, $e_1 \Rightarrow e_2$}$ \;
    \Return $(\pc_2, \sv{e_1} \wedge \sv{e_\Rightarrow})$ \;
\end{myalgo}
\vspace{-1.5em}
\caption{Additional symbolic execution rules for evaluating pure expressions.}
\label{fig:rules_eval_more}
\end{figure}

Viper's remaining symbolic execution rules for evaluating expressions
did not need to be changed when we implemented our technique.
For illustrative purposes, we show the rule for evaluating
heap-\emph{in}dependent functions (including arithmetic and other operators),
and for evaluating short-circuiting conjunction.

\newpage
%%--------------------------------------------------------------------
\section{Descriptions of Examples}
\label{sec:appendix_descriptions}
%%--------------------------------------------------------------------

\begin{itemize}
  \item \code{arraylist} %% vmcai2016/arraylist-quantified-permissions.sil
    is an encoding of a list implemented on top of an array, with operations to append an element to the list, and to insert an element into the list such that the list, if it was sorted before, remains sorted afterwards.

  \item \code{array-quickselect} %% quickselect/arrays_quickselect_rec.sil
    is an encoding of a (recursive) quickselect implementation over an array, with strong specifications such as ``the array has been permuted'', and ``the $n$-th smallest element has been selected''.

  \item \code{binary-search-array} %% binary-search/binary-search-array.sil
    is an encoding of an (iterative) binary search performed over a sorted array.

  \item \code{graph-copy} %% graph-copy/graph-copy.sil
    is the encoding of an algorithm that copies a graph. Its specifications make use of a custom axiomatisation of maps to record relations between original and copied nodes.

  \item \code{graph-marking} %% graph-marking/graph-marking.sil
    is the encoding of a graph marking algorithm, in the spirit of mark-and-sweep garbage collectors, with strong specifications such as ``nodes reachable from marked nodes are marked themselves''.

  \item \code{longest-common-prefix} %% longest-common-prefix\/ongest-common-prefix.sil
    is a challenge from the VerifyThis Verification Competition 2012: finding the longest common prefix of two arrays.

  \item \code{max-array-elimination} %% max_array/max-array-elimination.sil
    is a challenge from the COST Verification Competition 2011: finding the maximum in an array by elimination.

  \item \code{max-array-standard} %% max_array/max-array-standard.sil
    is an encoding of the straightforward way of finding the maximum in an array; it uses the same interface specifications and the same client as the previous example.

  \item \code{parallel-array-replace}
    is the running example from this paper: replace each occurrence of an element in an array segment by recursing over the two half-segments in parallel.
\end{itemize}

\section{Examples}
\label{sec:appendix_examples}
%----------------------------------------

\subsection{Running Example: Parallel Array-Replace}
\label{sec:appendix_replace}
%------------------------------------

\figref{code_replace} shows the encoding of our running example
(\inlinesilver{parallel-replace} from our test
set) in Viper.
Here, \inlinesilver{loc(a,i)}
is the injective function mapping an array \inlinesilver{a} to the ghost objects
modelling its array slots. So, a source-level expression \inlinesilver{a[i]} is
translated to \inlinesilver{loc(a,i).val} (see also \secref{inhale_exhale_permissions}).
Our code defines the pre- and postconditions of the \inlinesilver{Replace}
method as parameterised macros (occurrences of which are inlined,
similar to C-style macros), for reuse when encoding the recursive parallel calls.
Viper does not support parallel composition, but fork-join-style concurrency can be
modelled by appropriate \inlinesilver{exhale} (fork) and \inlinesilver{inhale} (join)
statements.

\begin{figure}[t]
\begin{silver}[mathescape=true]
define pre1(a, l, r) 0 <= l and l < r and r <= len(a)
define pre2(a, l, r)  forall i: Int :: l <= i and i < r ==>
                                            acc(loc(a, i).val)
define post1(a, l, r) forall i: Int :: l <= i and i < r ==>
                                            acc(loc(a, i).val)
define post2(a, l, r) forall i: Int :: l <= i and i < r ==>
                      (old(loc(a, i).val == from)
                         ? loc(a, i).val == to
                         : loc(a, i).val == old(loc(a, i).val))

method Replace(a: Array, left: Int, right: Int, from: Int, to: Int)
  requires pre1(a, left, right)
  requires pre2(a, left, right)
  ensures  post1(a, left, right)
  ensures  post2(a, left, right)
{
  if (right - left <= 1) {
    if(loc(a, left).val == from) {
      loc(a, left).val := to
    }
  } else {
    var mid: Int := left + (right - left) \ 2

    //fork-left
    exhale pre1(a, left, mid)
    exhale pre2(a, left, mid)

    //fork-right
    exhale pre1(a, mid, right)
    exhale pre2(a, mid, right)

    //join-left
    inhale post1(a, left, mid)
    inhale post2(a, left, mid)

    //join-right
    inhale post1(a, mid, right)
    inhale post2(a, mid, right)
  }
}
\end{silver}
\caption{Our running example, encoded in Viper. Inlined macros are used to
reuse the pre- and postcondition of \inlinesilver{Replace} when encoding
the parallel recursive calls.}
\label{fig:code_replace}
\end{figure}

\figref{code_array} shows the background definitions for the array encoding
that is used in \figref{code_replace} (as well as in other array-related
examples from our test suite).
Axiom \inlinesilver{all_diff} constrains function \inlinesilver{loc} to be
injective in both arguments by axiomatising \inlinesilver{first} and
\inlinesilver{second} to be the inverse functions for the first and second
parameter of \inlinesilver{loc}, respectively.

\begin{figure}[t]
\begin{silver}[mathescape=true]
field val: Int

domain Array {
  function loc(a: Array, i: Int): Ref
  function len(a: Array): Int
  function first(r: Ref): Array
  function second(r: Ref): Int

  axiom all_diff {
    forall a: Array, i: Int :: {loc(a, i)}
      first(loc(a, i)) == a and second(loc(a, i)) == i
  }

  axiom length_nonneg {
    forall a: Array :: len(a) >= 0
  }
}
\end{silver}
\caption{Background definitions for our array encoding. It
declares a type \inlinesilver{Array}, an injective function
\inlinesilver{loc} denoting the ghost object representing the array slot
at a given index, and a function
\inlinesilver{len} that denotes the length of an array.}
\label{fig:code_array}
\end{figure}

\figref{code_client} shows a client that uses \inlinesilver{Replace}, and
a heap-dependent boolean function \inlinesilver{Contains} that yields true if
an array contains a given value in the array prefix
\inlinesilver{[0..before)}. \inlinesilver{Contains} is intentionally left
abstract (i.e., it has no body) to demonstrate that the only way of reasoning
about the function is via function framing, which indeed allows us to prove
the final assertion.

\begin{figure}[t]
\begin{silver}[mathescape=true]
method Client(a: Array)
  requires 1 < len(a)
  requires forall i: Int ::
              0 <= i && i < len(a) ==> acc(loc(a, i).val)
  requires Contains(a, 5, 1)
{
  Replace(a, 1, len(a), 5, 7)
  assert Contains(a, 5, 1)  // Requires function framing
}

function Contains(a: Array, v: Int, before: Int): Bool
  requires 0 <= before && before <= len(a)
  requires forall i: Int ::
              0 <= i && i < before ==> acc(loc(a, i).val)
\end{silver}
\caption{Client of the \inlinesilver{Replace} method from
\figref{code_replace}. Function framing allows us to prove the assertion in
method \inlinesilver{Client}.}
\label{fig:code_client}
\end{figure}

% \afterpage{\clearpage}
\subsection{Graph-Marking}
\label{sec:appendix_graph}
%------------------------------------

\figref{code_graph} shows an encoding of a graph-marking algorithm
(\inlinesilver{graph-marking} from our test set)
in Viper. In Viper, the double ampersand (\inlinesilver{&&}) is overloaded:
it denotes the separating conjunction ($\ast$) as well as the usual boolean
conjunction ($\wedge$); in the conjunction of two impure assertions, it always
denotes the separating conjunction.

The macro \inlinesilver{INV} describes a graph in terms
of accessibility predicates and closure properties over a given set of
nodes (of the graph): the first three \inlinesilver{forall}s are \iscs,
denoting permissions to the fields of each node in the set of nodes.
The remaining two \inlinesilver{forall}s are pure quantifiers; they
express that the set of nodes is closed under following the \inlinesilver{left}
and \inlinesilver{right} fields.
The two quantifiers have been annotated with triggers to improve performance,
as is common for Viper encodings.

\begin{figure}[t]
\begin{silver}[mathescape=true]
field left: Ref;  field right: Ref;  field marked: Bool

define INV(nodes)
     !(null in nodes)
  && (forall n: Ref :: n in nodes ==> acc(n.left))
  && (forall n: Ref :: n in nodes ==> acc(n.right))
  && (forall n: Ref :: n in nodes ==> acc(n.marked))
  && (forall n: Ref :: {n.left in nodes}{n in nodes, n.left}
        n in nodes && n.left  != null ==> n.left  in nodes)
  && (forall n: Ref :: {n.right in nodes}{n in nodes, n.right}
        n in nodes && n.right != null ==> n.right in nodes)

method trav_rec(nodes: Set[Ref], node: Ref)
  requires node in nodes && INV(nodes) && !node.marked
  ensures node in nodes && INV(nodes)
  /* Marked nodes are not unmarked */
  ensures forall n: Ref :: {n in nodes, n.marked}
            n in nodes ==> (old(n.marked) ==> n.marked)
  ensures node.marked
  /* The graph structure is not modified. */
  ensures forall n: Ref :: {n in nodes, n.left}
            n in nodes ==> (n.left == old(n.left))
  ensures forall n: Ref :: {n in nodes, n.right}
            n in nodes ==> (n.right == old(n.right))
  /* Propagation of the marker */
  ensures forall n: Ref :: {n in nodes, n.marked}
                           {n in nodes, n.left.marked}
            n in nodes ==>
              (  old(!n.marked)
               && n.marked ==> (n.left == null || n.left.marked))
  ensures forall n: Ref :: {n in nodes, n.marked}
                           {n in nodes, n.right.marked}
            n in nodes ==>
              (   old(!n.marked)
               && n.marked ==> (n.right == null || n.right.marked))
{
  node.marked := true

  if (node.left != null && !node.left.marked) {
    trav_rec(nodes, node.left)
  }

  if (node.right != null && !node.right.marked) {
    trav_rec(nodes, node.right)
  }
}
\end{silver}
\caption{An encoding of a simple graph-marking algorithm in Viper.}
\label{fig:code_graph}
\end{figure}

\end{appendix}

\end{document}

%% file: macros.tex
\usepackage[normalem]{ulem}
 \newcommand{\asout}[1]{}

\newcommand*{\term}[1]{\textit{#1}}

\newcommand*{\figref}[1]{Fig.~\ref{fig:#1}}
\newcommand*{\secref}[1]{Sec.~\ref{sec:#1}}
\newcommand*{\appref}[1]{App.~\ref{sec:#1}}
\newcommand*{\Isc}{ISC\xspace}
\newcommand*{\isc}{ISC\xspace}
\newcommand*{\Iscs}{\Isc{s}\xspace}
\newcommand*{\iscs}{\isc{s}\xspace}
\newcommand*{\ic}{quantified chunk\xspace}
\newcommand*{\ics}{\ic{s}\xspace}
\newcommand*{\FVF}{value map\xspace}
\newcommand*{\FVFs}{value maps\xspace}
\newcommand*{\naive}{na\"{\i}ve\xspace}

\newcommand*{\reftype}{\mathit{Ref}}

\DeclareMathOperator*{\fvftype}{\mathit{PVM}}
\DeclareMathOperator*{\snaptype}{\mathit{Snap}}
\newcommand*{\settype}[1]{\mathit{Set}[#1]}
\DeclareMathOperator*{\fvf}{\mathit{v}}
\DeclareMathOperator*{\pvs}{\mathit{pvm}}
\DeclareMathOperator*{\fun}{\mathit{fun}}

\newcommand*{\hp}{\ensuremath{h}\xspace}
\newcommand*{\pc}{\ensuremath{\pi}\xspace}

\newcommand*{\triggers}[1]{\ensuremath{\textbf{\{}#1\textbf{\}}}\xspace}
\newcommand*{\VmDefEq}{\textsc{VmDefEq}\xspace}

\ifx\symlasy\undefined   \DeclareSymbolFont{lasy}{U}{lasy}{m}{n}
  \SetSymbolFont{lasy}{bold}{U}{lasy}{b}{n} \else
\fi
\DeclareMathSymbol\safeleadsto {\mathrel}{lasy}{"3B}

\newcommand*{\sil}[1]{\ifmmode\mbox{\inlinesilver{#1}}\else\inlinesilver{#1}\fi}

\newcommand*{\pointsto}[2]{\ensuremath{{#1}\mapsto{#2}}}
\newcommand*{\inv}[1]{{\operator{\ensuremath{{#1}^{-1}}}}}
\newcommand*{\inve}{\inv{e}}
\newcommand*{\svinve}{\operator{\ensuremath{e^{-1}}}}
\newcommand*{\ite}[3]{\ensuremath{#1\ ?\ #2\, :\, #3}}

\newcommand*{\sv}[1]{\ensuremath{\underline{#1}}}
\newcommand*{\inva}{\operator{\ensuremath{\sv{a}^{-1}}}}
\newcommand*{\compilesto}{\safeleadsto}

\newcommand{\eg}{{\it{e.g.\@}}}

\newcommand{\etal}{{\it{et al.\@}}}

\newenvironment{myalgo}[1][ht]{%
\RestyleAlgo{boxed}
\SetAlCapSkip{0.75em}
\SetCommentSty{}
\begin{algorithm}[#1]
\AlgoDontDisplayBlockMarkers%
\SetAlgoNoLine%
\DontPrintSemicolon%
}{%
\end{algorithm}
}

\newcommand*{\algoname}[1]{#1}
\SetKwFunction{funSplitQP}{\algoname{remove}}
\SetKwFunction{funValueQP}{\algoname{summarise}}
\SetKwFunction{funPermQP}{\algoname{perm}}

\SetKwProg{Fn}{def}{\string:}{}
\SetKwIF{If}{ElseIf}{Else}{if}{then:}{else if}{else:}{}%
\SetKwFor{For}{foreach}{do:}{end}%
\SetKw{Var}{var}
\SetKw{Check}{check}
\SetKwFunction{Exhale}{exhale}
\SetKwFunction{Inhale}{inhale}
\SetKwFunction{Assume}{assume}
\SetKwFunction{Assert}{assert}
\SetKwFunction{Fail}{fail}
\SetKwFunction{Eval}{eval}
\SetKwFunction{EvalExt}{eval'}
\SetKwFunction{Exec}{exec}

\renewcommand*{\gets}{~{:=}~}%
\renewcommand*{\;}{\\}%

\newcommand*{\symb}[1]{\ensuremath{\mathit{#1}}}
\newcommand*{\func}[1]{\ensuremath{\mathit{#1}}}
\newcommand*{\permmin}{\symb{min}}

\newcommand*{\eqs}{\symb{def}}
\newcommand*{\perms}{\symb{perm}}

\newcommand*{\pt}{q_{\symb{current}}}
\newcommand*{\ptt}{q_{\symb{needed}}}

\newcommand*{\domain}{\func{domain}_f}
\newcommand*{\lookup}{\func{apply}_f}
\newcommand*{\toSnap}{\func{toSnap}}

\newcommand*{\FV}{\operator{\ensuremath{\mathit{FV}}}}

\def\vs{\fvf}

\newcommand{\sub}[3]{\ensuremath{#1[#2/#3]}}

\newcommand{\hlLine}[1]{\leavevmode\rlap{\hbox to \hsize{\color{Lightgray}\leaders\hrule height .8\baselineskip depth .9ex\hfill}}#1}
\newsavebox{\newBox}
\newlength{\newWidth}
\newcommand{\hl}[1]{
\savebox{\newBox}{\ensuremath{#1}}\settowidth{\newWidth}{\usebox{\newBox}}\leavevmode\rlap{\hbox to \hsize{\color{Lightgray}\leaders\hrule height .8\baselineskip depth .9ex\hskip \newWidth}}\usebox{\newBox}}

\newcommand*{\sforall}{\code{forall}\xspace}
\newcommand*{\snone}{0\xspace}
\newcommand*{\swrite}{1\xspace}
\newcommand*{\qpforall}[3]{\textup{\sforall \code`#1` \code`::` \code`#2` \ensuremath{\Rightarrow} \code`#3`}}
\newcommand*{\chunk}[4]{\ensuremath{\pointsto{#1.#2}{[#3(#1), #4]}}}

\makeatletter
\def\operator#1{\@ifnextchar\bgroup {\operatorarg{\ensuremath{#1}}}{\ensuremath{#1}}}
\def\operatorarg#1#2{{#1}{\ensuremath{(#2)}}}
\def\spoperator#1{\@ifnextchar\bgroup{\spoperatorarg{\ensuremath{#1}}}{\ensuremath{#1}}}
\def\spoperatorarg#1#2{\ensuremath{#1~#2}}

\def\fixedoperator#1{\@ifnextchar\bgroup {\fixedoperatorarg{#1}}{\ensuremath{#1}}}
\def\fixedoperatorarg#1#2{\fixedoperatorparse{#1}#2~}
\def\fixedoperatorparse#1#2,#3~{\ensuremath{{#2}{.}{#1}{(#3)}}}
\makeatother